Nikolaos Polatidis

Supervisor: Dr Peter Hancox

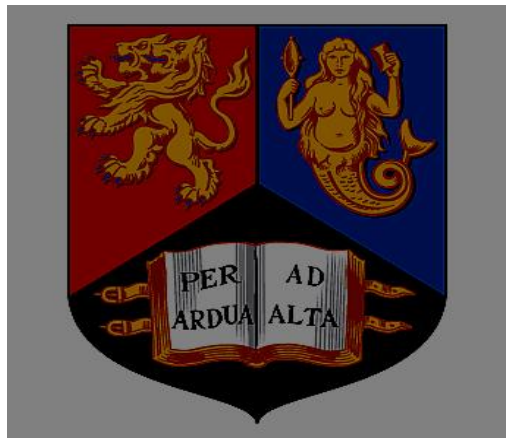

# Chatbot for admissions

Submitted for the degree of

MSc in Internet Software Systems

School of Computer Science

University of Birmingham

# Abstract

The communication of potential students with a university department is performed manually and it is a very time consuming procedure. The opportunity to communicate with on a one-to-one basis is highly valued. However with many hundreds of applications each year, one-to-one conversations are not feasible in most cases. The communication will require a member of academic staff to expend several hours to find suitable answers and contact each student. It would be useful to reduce his costs and time.

The project aims to reduce the burden on the head of admissions, and potentially other users, by developing a convincing chatbot. A suitable algorithm must be devised to search through the set of data and find a potential answer. The program then replies to the user and provides a relevant web link if the user is not satisfied by the answer. Furthermore a web interface is provided for both users and an administrator.

The achievements of the project can be summarised as follows. To prepare the background of the project a literature review was undertaken, together with an investigation of existing tools, and consultation with the head of admissions. The requirements of the system were established and a range of algorithms and tools were investigated, including keyword and template matching. An algorithm that combines keyword matching with string similarity has been developed. A usable system using the proposed algorithm has been implemented. The system was evaluated by keeping logs of questions and answers and by feedback received by potential students that used it.

# Acknowledgements

I would like to thank Dr Peter Hancox for his immeasurable help and support throughout this project. I also need to express my thanks to the computer support team for their excellent help and instructions.

Finally, I feel the need to acknowledge the constant support offered by my parents.

# Table of Contents

# Figures

# Tables

# Chapter 1

# Introduction

This chapter covers an introduction to the project including the context, a description of aims and objectives, a description of what has been achieved, contributions and the structure of the report.

## 1.1 Context

Although the admissions process works properly as it is, it is very difficult and time consuming to contact a member of staff of the university. However, the problem would be partially solved if the applicant could talk to a convincing chatbot, able to respond to their concerns with information about admissions, booking accommodation, paying fees in instalments and what pre-sessional courses are on offer.

The chatbot should be able to communicate with a user in a way similar to the following:

ChatBot: Hello how can I help you?
User: What is the minimum score of IELTS required for entry into MSc Computer Science?
ChatBot: The minimum requirement is a total score of 6.0 with no less that 5.5 in any section.
User: What other test is accepted by the department?
ChatBot: We also accept TOEFL and the certificate of Proficiency of the University of Cambridge.
User: What are the entry requirements for the MSc in Computer Security?
ChatBot: To enter the programme, you need to have at least an Upper Second Class (2.1) degree or an international equivalent in Computer Science or a closely related discipline and a solid foundation in programming.

## 1.2 Aims and Objectives

The aim of this project is to contribute to the solution of the problem of direct communication between applicants and the university.

The main objectives of the project are as follows:

**Database:** To develop a database were all the relevant information about questions, answers, keywords, logs and feedback will be stored.

**Algorithm:** To develop a keyword matching algorithm and a string distance comparison algorithm and combine them in order to retrieve the best possible answer.

**Interface:** To develop a web interface which aims to give the ability to potential students and their families to submit questions in a chatbot and get convincing replies. The web system can be hosted on any computer that supports an operating system on which the Java Virtual machine will be able to load. Furthermore the Hibernate, Spring, Apache Wicket, Lingpipe, the Jazzy spelling, the Stanford NLP parser and Google search libraries should be available.

## 1.3 Achievements

The following goals were achieved:

- Investigated a range of algorithms for solving the problem of direct communication with the university and developed an algorithm that combines the use of keyword matching with string distance.

- Designed and implemented a usable chat system, which has been extended after being evaluated by users.

- Evaluated the extended system with users. The system was used by potential students of the university.

- Completed a literature research in order to learn more about natural language processing, investigated existing tools and established the requirements.

## 1.4 Methodology

The design and development of the system followed the waterfall model as described below.

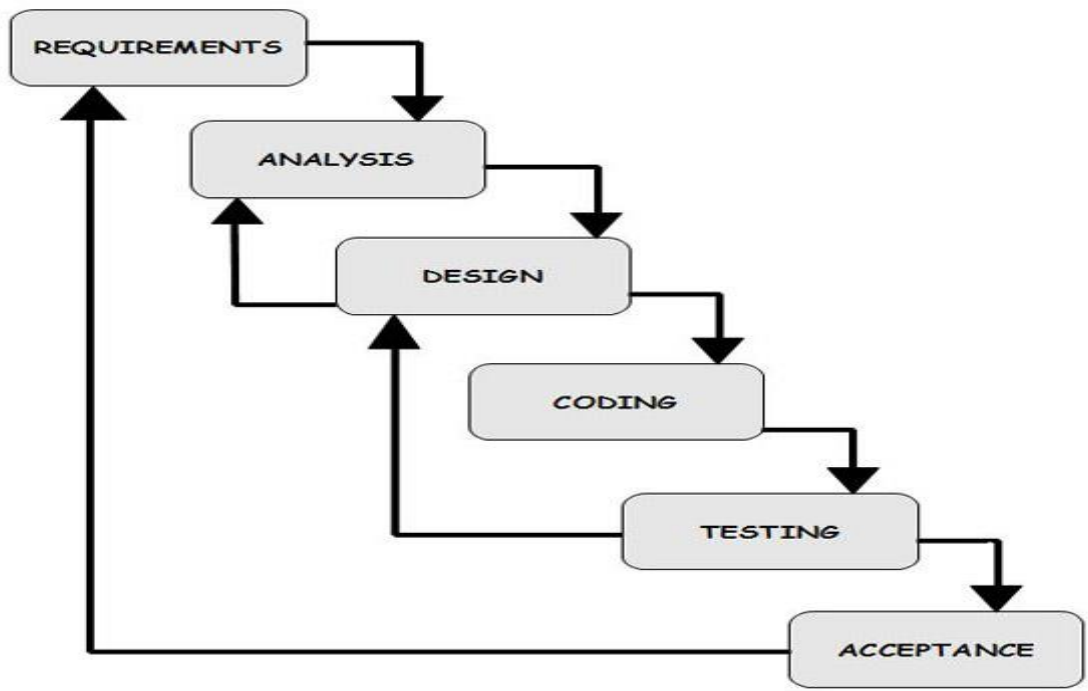

**Figure 1.1** Waterfall model

The waterfall model follows a series of processes, which are used during development. Usually the stages will require the gathering of requirements and their analysis. The design of the system is the next stage, followed by coding the actual system. Then evaluation, testing and debugging, if necessary, is the next step. Finally the system will either be accepted and therefore maintained or rejected. It is vital to move to the next process of the waterfall model if the previous step has been completed.

## 1.5 Contributions

A number of novel concepts have been incorporated to the system. A keyword matching algorithm has been used in combination with a string similarity algorithm to provide a better reply and is described in sections 4.2.1, 4.2.2 and 4.2.3. Also the Stanford parser has been used to analyse the structure of the given sentence as described in section 4.2.5. Finally the system has been evaluated using feedback and logs from potential students that used it, described in section 6.2.

## 1.6 Outline of Dissertation

This report is divided into several chapters.

Chapter two covers the background research, which gives an overview of chatbots, interactions of chatter-bots with humans, matching algorithms and current systems. Chapter three gives and overview and covers the requirements of the system. Chapter four covers the design of the system. It includes the database design, an explanation of the algorithms and how they work. It goes on to explain the functionality provided by the system for the users and the administrator. Chapter five covers the implementation of the system. The system is described with further details on the technologies used. Chapter six contains the evaluation and testing of the final system. Chapter seven is the conclusion of the project. It contains a summary, the main achievements, its limitations and future work.

# Chapter 2

# Background Research

This chapter contains the background research about topics relevant to the admissions communication. It presents an overview of chatbots and their interaction with humans. Furthermore the keyword matching and the template matching algorithms are mentioned, followed by real world examples of their use.

## 2.1 Chatbots

A chatbot is a program that is used to participate in conversations with humans. It uses an appropriate interface for input and output and with the use of AI techniques it can provide realistic answers so the user will think that the communication taking place is with another human. The implementation of such systems varies from using keyword matching, string similarity or complex natural language processing techniques. More sophisticated chatbots could learn from the user input. Nowadays chat bots are used widely in web applications in order to provide help or information when it is asked by the users.

*"Chatbots are computer programs that interact with users using natural languages*[21]."

## 2.2 Existing applications

Chatbots are used in many organisations instead of humans when this is possible. Most of the time systems like that are based on the ways ELIZA or ALICE communicates.

### 2.2.1 ELIZA

The first chatbot developed was ELIZA [21]. It was developed by Joseph Weizenbaum using a keyword matching technique. The idea was to read the user input and search for certain keywords, if a keyword was found then the answer was retrieved. If a keyword was not present then ELIZA would try, according to specified rules, to get more information from the user to keep the conversation going. Consider the following example to understand how this would work.

User: How can I get in touch with my mother?

Then the chatbot would identify the keyword mother and would find a connection between mother and family and it would respond with a question.

ELIZA: Tell me more about your family

The chatbot doesn't understand what it is saying. It only produces results according to rules and in this case there was a rule connecting mother and family. Then it had to produce an answer that would encourage the user to submit another question with the hope of finding a keyword. However most of the time there are some standard replies that ELIZA would use if a keyword is not found. Some examples would be "Very interesting, please go on" or "Can you think of a special example?"

The following figure gives a good example of chatting with ELIZA.

```
Human:  Men are all alike.
ELIZA: In what way?
Human:  They're always bugging us about something.
ELIZA: Can you think of a special example?
Human:  Well, my boyfriend made me come here.
ELIZA: Your boyfriend made you come here?
Human:  He says I'm depressed much of the time.
ELIZA: I'm sorry to hear that you are depressed much of the time.
Human:  It's true.  I'm unhappy.
ELIZA: Do you think coming here will help you not to be unhappy?
Human:  I need some help, that much seems certain.
```

**Figure 2.1** A sample conversation with ELIZA (Jurafsky and Martin, 2000)

### 2.2.2 A.L.I.C.E

ALICE was implemented by Richard Wallace in 1995[21]. It uses pattern matching and stores the information in Artificial Intelligence Mark-up Language, or else known as AIML, files. An AIML file is similar to an XML file that was developed to store pattern knowledge for chatbots. There are three types of AIML categories, atomic categories, default categories and recursive categories, which are described below.

Atomic categories: Is a type of AIML category where there is an exact match.

<category>
        <pattern>What is your name</pattern>
        <template>My name is Nick</template>
</category>

In the above example if the user submits "What is your name" then the chatbot will reply with "My name is Nick".

Default categories: Is a type of AIML category where the use of a wild character such as * is used to match any input.

<category>
        <pattern>What is *</pattern>
        <template>It is my name</template>
</category>

If ALICE does not find a solution using the first category it will proceed to the second. In this case if a user submits "What is a name" then ALICE will do a check until the wild char and if there is a match it will come with the answer.

Recursive categories: Is a type of an AIML category where special tags are used to refer to a recursion to convince the user to be more specific.

```
<category>
        <pattern>Do you know what the * is</pattern>
        <template>It is my name
        <srai>What is <star/></srai>
</template>
</category>
```

In this case the wild character * is used to reduce the input submitted by the user. If the user submits "Do you know what the time is" then this becomes "What is the time".

The following figure gives an example of a conversation between a human and ALICE.

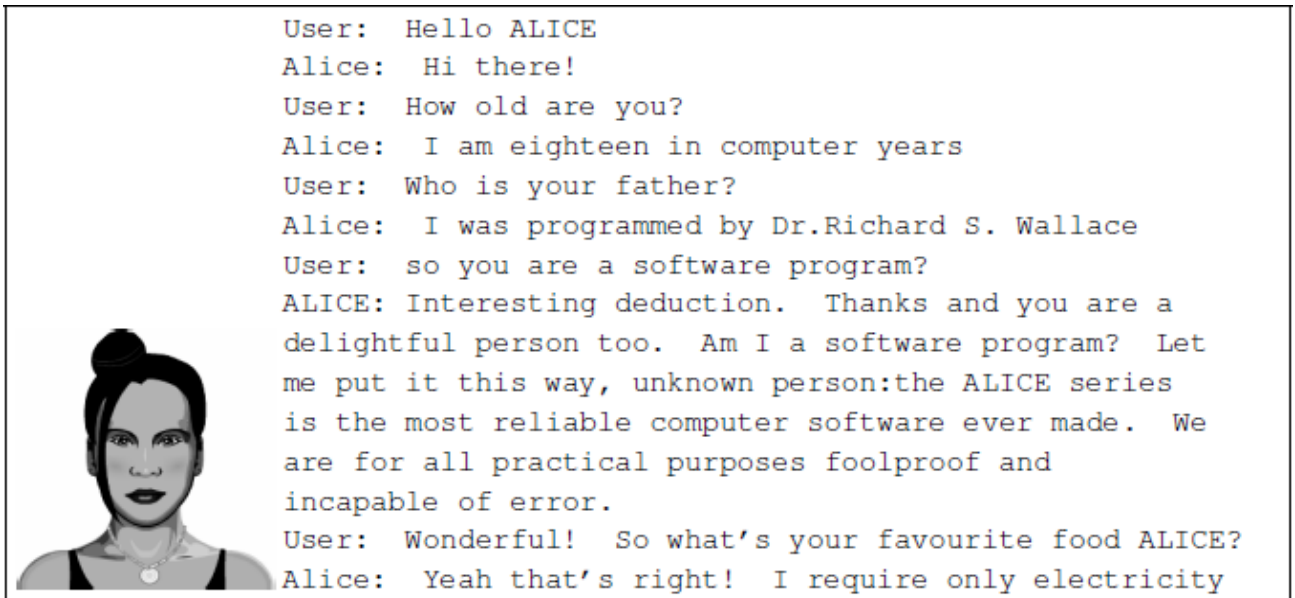

**Figure 2.2** A sample conversation with ALICE (Shawar and Atwell, 2007)

When a user submits a question then any relevant punctuation is removed and the sentence goes through the categories to find the longest pattern match. The question "Can you tell me the time, using hours, minutes and seconds?" will become» Can you tell me the time using hours minutes and seconds" and then it will search to find the longest match. When the match is found then it will be forwarded to the output. However ALICE does not have the ability to learn and can only come with data that exist in its database.

## 2.3 Learning chatbots

Learning chatbots have the ability to learn from user input. The intelligence of the chatbot is based on specified rules and by assuming that the user will give a correct answer. There are two types of learning and both use the same learning rules. The first one will understand what the user said, will store it for the current session and will reply with that answer if the user asks about. The second one will store the information in its database and will reply with that answer for any user. Certain rules can be specified to understand from previous user input if the possibility of a user saying the truth is high or not and decide to store the information for future users. Also questions and answers among users can be saved in the database and compared to identify the possibility of something being true.

The following conversation gives an example of the learning procedure.

User: How many lecturers are there in the department of computer science?
Chatbot: I am sorry but I don't know the answer. Perhaps you can help me learn
User: There are 28 lecturers at the department
Chatbot: Is this the correct answer?
User: Yes
Chatbot: Thank you for helping
User: How many lecturers are there in the department of computer science?
Chatbot: There are 28 lecturers at the department

## 2.4 Interaction with humans

Humans believe that the concept of a chatbot is not realistic and that a chatbot will not be able to provide satisfying answers. Most of the time chatbots are being developed to serve certain purposes and if a human tries to trick the chatbot it is likely that he will succeed. Moreover humans do not think that a computer program will be able to satisfy their needs and they do not even try to interact. Also if a human is aware that a chatbot has extensive knowledge then he will be more demanding. In the case that we have to convince a human that the conversation is taking place with another human then an extensive database has to be used along with certain rules such as answering multiple questions, occurrence of words and sentence parsing.

The Loebner prize is a chatbot competition with a number of judges communicating with both humans and chatbots at the same time and is based on a standard Turing test. There are a number of rounds in which a human judge is communicating with a chatbot and with a human that use a computer interface. Based on the answers the judge will have to decide which one is the chatbot and which one is the human, within a five minute limit.

Anaphora resolution is a very important aspect for a human like chatbot. It is a way of finding to what an expression refers.

*We found a room under the university accommodation, but it didn't have an en-suite bathroom.*

It is clear for a human that the bathroom refers to the room. However it would be more difficult for a chatbot to understand. However there are cases that an expression can become complex and therefore difficult to understand.

*I accepted my offer to the university because it was unconditional.*

*I accepted my offer to the university because it was one of the top.*

In the first case the expression refers to the offer, whereas in the second case refers to the university.

## 2.5 Web applications

The classic architecture is the three tier architecture as described in Figure 2.3. The advantages of using this type of model is that the functionality is distributed across three independent systems and also that any of the tier can be replaced or upgraded independently if the requirements or the technology change.

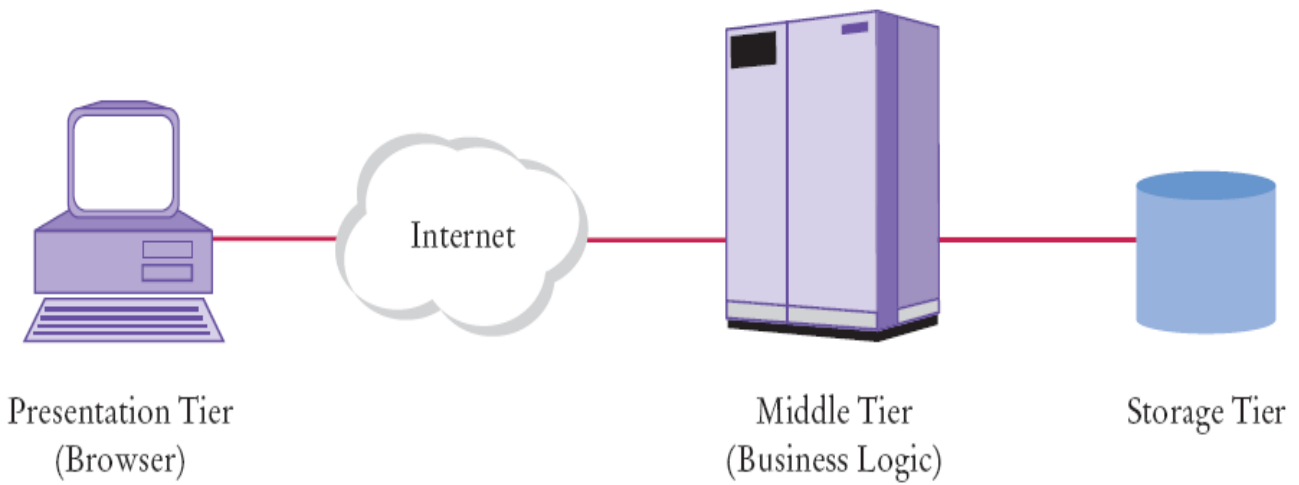

**Figure 2.3** Three Tier Architecture

## 2.6 Web application technologies

A number of technologies and libraries were investigated for the purpose of development of this project and the following were identified as the ones to be used for creating the system.

### 2.6.1 Java

Java is an object oriented programming language that can be used to develop software that it can run on a variety of operating systems without modification. Moreover it is simple to use language, with a large number of APIs, sufficient documentation and it is free.

### 2.6.2 PostgreSQL

PostgreSQL is an open source relational database management system. It has been chosen because it is supported by a variety of operating systems, it works efficiently with increasing load and has many advanced features. Furthermore there are a number of command line, web and GUI front end tools that can be used to manipulate databases.

### 2.6.3 Hibernate

Hibernate is an open source object-relational mapping set of APIs for the Java language. The purpose of Hibernate is to map the problem of mismatch between relational database systems and object. The idea is to have a Java class for each table and manipulate the objects created by the classes that map the tables. Moreover its thorough documentation and the fact that it is lightweight make it a wise choice.

### 2.6.4 Spring

Spring is a lightweight open source application framework that can be easily used with other frameworks. It has a number of features that make it essential to use. There is a centralized configuration file where Java beans can be created, transactions can be handled, without having to manage each transaction separately and security can be managed. Data access objects can be used on top of the Hibernate, which reduces errors and simplifies the development process. Testing application code written using Spring is easier than other similar technologies.

### 2.6.5 Apache Wicket

Apache Wicket is a Java, component based, web application framework that is easy to integrate with Spring and Hibernate. It provides a clear separation between static HTML and dynamic Java parts of the presentation layer. It allows the creation of reusable components. Wicket provides authorization roles that can be easily placed on top of each page. Also errors are identified during the compilation process.

### 2.6.6 Javascript

Javascript is a scripting, client-side, programming language. It can be used to provide encase the functionality of HTML pages. It can be used for various purposes such as setting character limitations on text areas.

### 2.6.7 HTML and CSS

HTML is the standard mark-up language used for the development of web interfaces. It must be used in combination with CSS, which is responsible for the layout of the HTML components.

The CSS files have been based on the CSS files developed by Mike Stenhouse[12].

### 2.6.8 Third party libraries

The Jazzy spelling Java API has been used and requires an interaction with a local dictionary file in order to provide a spelling correction if necessary. The Google search Java API has been used and requires communication with the Google search server on user request. Also the lingpipe Java natural language processing API has been used for calculating the string similarity. The Stanford natural language parser has been used for the analysis of a sentence and validity checking.

## 2.7 Other ideas

A combination of the keyword matching algorithm with the string similarity algorithm should be used. The keyword matching will start identifying keywords and retrieve the answer that has the most keywords. In the case that the same number of keywords is found in more than one answer or there are no keywords then the String similarity algorithm will run to identify the which question is most similar to the one entered and retrieve the answer.

## 2.8 Conclusion

From the information gathered from the research the strongest points and weaknesses of each have been identified and a list of desired features that should be in the system has been created. Finally the conclusion was that the system should contain a combination of the keyword matching with the String similarity algorithm, a customized google search option and a spell checker, log and feedback functionality. Furthermore an administration part should be created where the administrator will be able to add, update and delete information related to the system.

# Chapter 3

# Analysis and Specification

This chapter contains an overview, a system proposal and a thorough analysis of functional, non-functional requirements and users of the system.

## 3.1 Overview

The role of the system is to provide a chatbot that will be able to answer questions related to the admission procedure. It will provide a web interface for the users to interact with the system and an administration interface.

A user is anyone who would like to visit the website and engage in a conversation. As well as talking the user should be able to submit a log of whether he is satisfied with an answer and produce a link. Other than communicating, the user should be able to rank the system. The rank of the system should be a five star ranking system where one is poor and 5 is excellent. A user should also be able to write a review using the feedback form.

The administrator of the system shall be able to log in using a user name and a password. The responsibility of the administrator will be to maintain the system by adding questions and answers to the database and by updating current information sets when necessary. Furthermore he should be able to view the user ranking, feedback messages and logs.

The system shall provide its users with spell checking suggestions on screen, when they make such errors. Moreover the parsing of sentences will avoid sending to the system words that do not form a sentence.

There are several security issues which need to be taken into consideration when designing the system. These include personal and sensitive information. The data should not be accessed without authorisation and authentication. Disclosure or leak of data should be protected against various types of attacks and be encrypted and password protected.

## 3.2 Functional Requirements

1. *Chatting*:

   a. The system should allow users to chat.
   b. The system shall inform the user if an answer is not available.
   c. The system shall inform the user about spelling mistakes.
   d. The system shall inform the user about the validity of the sentence.

2. *Searching:*

   a. The system should allow users to search for information about admissions.
   b. The system should allow users to search for information about tuition fees.
   c. The system should allow users to search for information about accommodation.

3. *Logs:*

    a. The system should maintain a log of the current question and answer if the user is not satisfied.

4. *Feedback:*

    a. The user should be able to leave feedback, which is comprised of a text message and a rating.

5. *Administrative system*

    a. *Information management:* The administrator should be able to to add, update and delete questions, answers and keywords.

    b. *Log management:* The administrator should be able to view and delete logs.

    c. *Feedback management:* The administrator should be able to view and delete feedbacks.

## 3.3 Non-Functional Requirements

1. *User Interface:*

    a. The system shall maintain an easy to use interface across all functionality and for all users
    b. The clients' user interface should be compatible with all commonly used browsers, such as Internet explorer, Firefox, Google chrome and Safari.

2. *Scalability:*

    a. The system shall be able to scale based on the number of users using the system.

3. *Security:*

    a. The administrative system should be protected from unauthorized access.
    b. The database should protected from attacks and unauthorized access.
    c. The interface should be protected from attacks.
    d. All passwords should be stored as a secure hash of the administrator password.

4. *Third party interactions:*

    a. The system should be able to interact with the Google spelling server, which handles the spelling.
    b. The system should be able to interact with the Google search server, which is used for the customized search on the admissions website.

5. *Portability:*

    a. The system should run on a variety of operating systems that support the Java language.
    b. The system should run on a variety of hardware.

6.  *Maintainability:*

   a. The system should be easy to maintain.
   b. There should be a clear separation of HTML and Java interface code.
   c. There should be a clear separation between the interface and the business logic code.
   d. There should be a clear separation between the data access objects that map the database and the business logic code.

7.  *Exception handling:*

   a. Exceptions should be reported effectively to the user if they occur.

8.  *Ethics:*

   a. The system shall not store or process  any information about its users.

## 3.4 Use case model

The use case diagram describes the functionality of the system as designed from the requirements and can be found below.

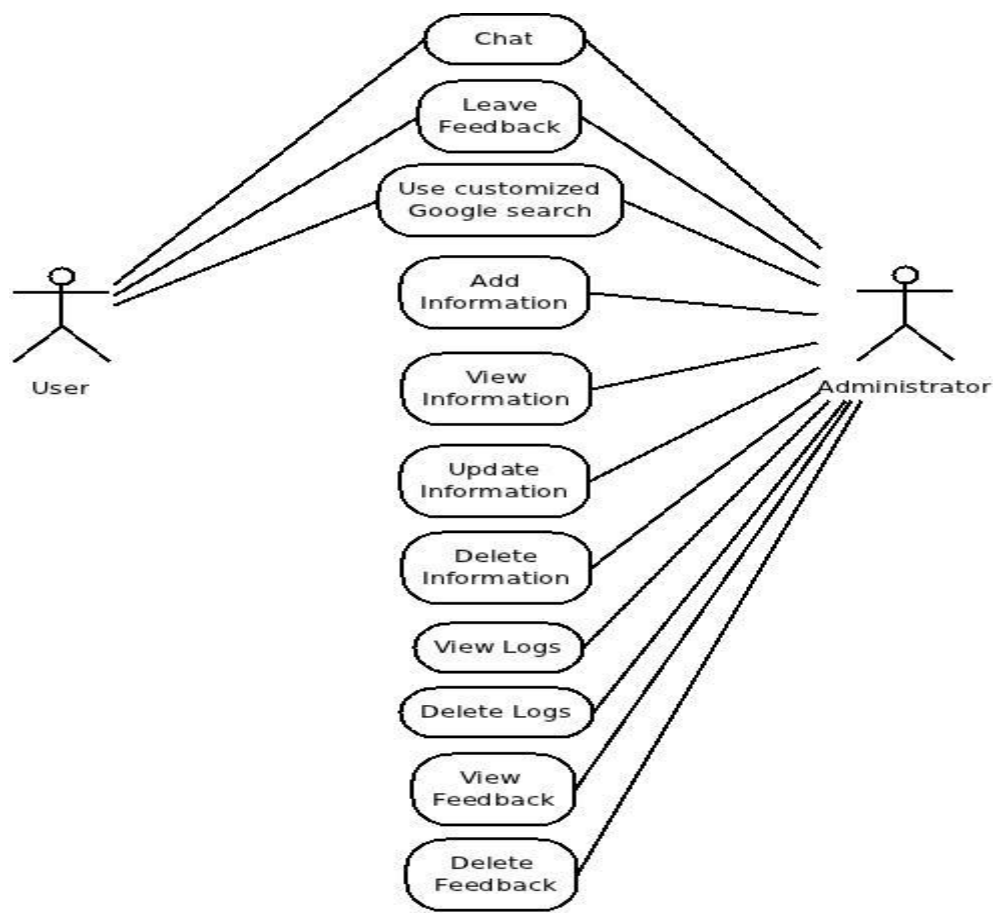

**Figure 3.1** Use case diagram

### 3.4.1 Actor documentation

| Actor | Description |
|---|---|
| User | Someone who uses the system to engage in a conversation |
| Administrator | The administrator of the system. Allowed to carry out administrative tasks |

**Table 3.1** Actor documentation

# Chapter 4

# Design

This chapter outlines the design of the system. It comprises a database, a combination of two algorithms, the business logic and the web interface. The system includes interfaces for users and an administrator.

## 4.1 Database design

The design of the database aims to keep it as simple and efficient as possible. The *user_roles* table will hold the user name and a role name, the *roles* table will store a description of each role and an id, the *chatbotuser* table will store the user details, the *information* table will store the questions, answers and keywords, the *fdback* table will store the feedback received and the *log* table will store the logs.

The figure below shows the entity relationship diagram designed for the database of the system.

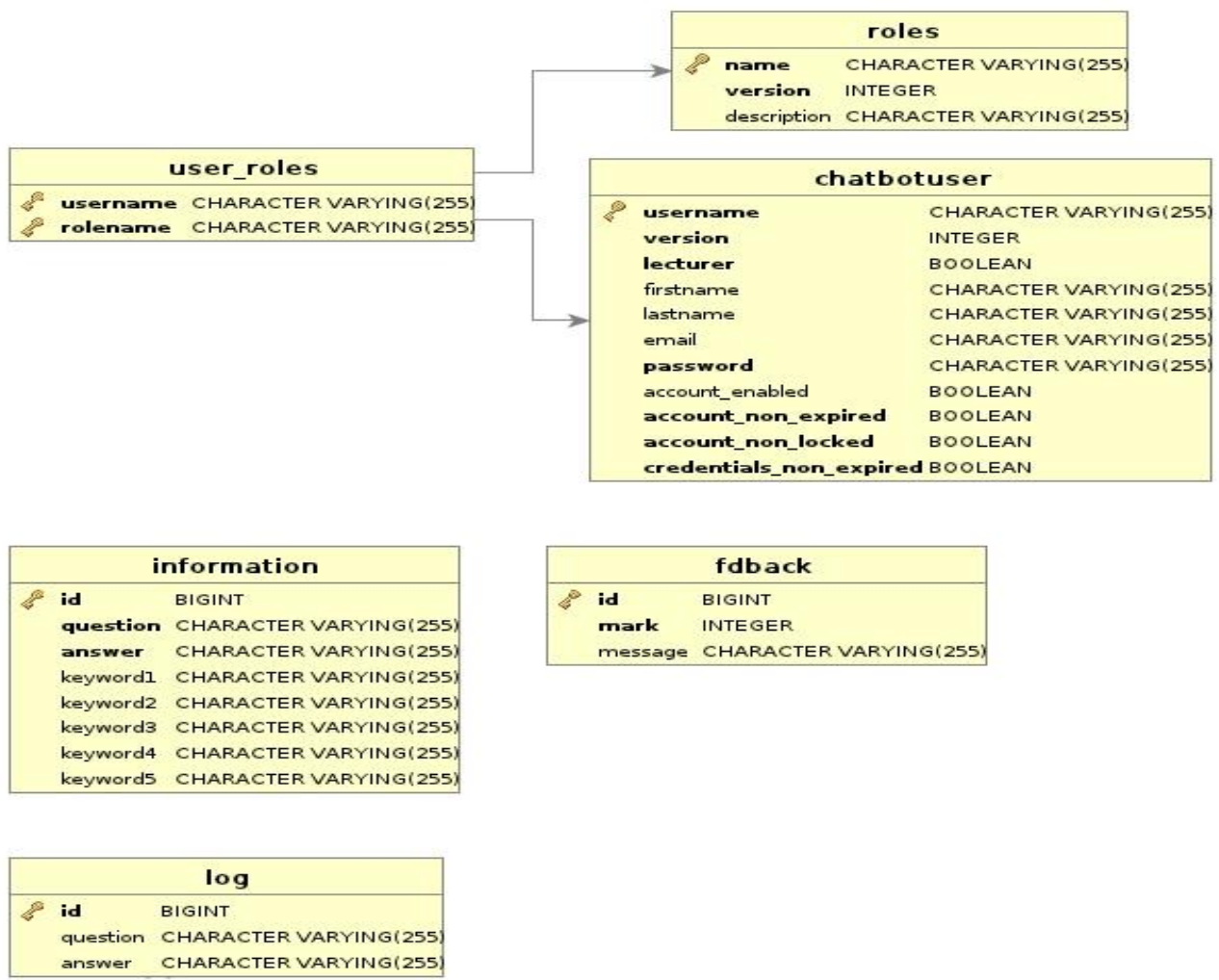

**Figure 4.1** Entity Relationship diagram

## 4.2 Algorithms

### 4.2.1 Keyword matching

The keyword matching algorithm will attempt to identify keywords in a sentence. In the case that one or more keywords are found in the user's input text then an answer will be retrieved. However, in the case that the same number of found keywords exist as keys for more than one answer then there is a deadlock. Again a deadlock occurs in the case that there are no keywords.

Consider the following example in order to understand better how the keyword matching algorithm works. The table represents a small number of information rows from the database of the system.

| Question | Answer | Keyword 1 | Keyword 2 |
|---|---|---|---|
| Do I need a visa? | Yes, unless you are from an EU/EEA country | Visa | |
| What are the entry requirements? | You need at least a 2.1 degree in order to be considered for a place in one of our courses | Entry | Requirements |

**Table 4.1** Keyword matching

The user then submits the question: "Does a 2.2 in computer science satisfy the entry requirements for the MSc in computer security?"

The algorithm will identify both keywords in the second row and will retrieve the answer.

### 4.2.2 String similarity

The string similarity algorithm will try to find similarities between a list of strings and will retrieve the answer that is the closest match to the question submitted by the user.

Consider the following example in order to understand better how the string similarity algorithm works. The table represents a small number of information rows from the database of the system.

| Question | Answer | Keyword 1 | Keyword 2 |
|---|---|---|---|
| Do I need a visa? | Yes, unless you are from an EU/EEA country | Visa | |
| Can I pay with a visa? | Yes, both tuition and accommodation fees can be paid online using a visa | Visa | |

**Table 4.2** String similarity

The user then submits the question: "Can I pay using visa?"

The algorithm will then compare the question of the user with the questions in the system and will respond with a percentage for each, ignoring the keywords, which represents the possibility of similarity between strings. In this case the second question has the closer match with the user submitted question and the answer from that information line will be retrieved.

The Jaro-Winkler distance has been used in the development of the algorithm[17]. The figure below gives the code walk through for the Jaro-Winkler distance.

**Code Walk Through**
The code is trivial. We simply use the distance constant in the Jaro-Winkler class, and then read in arguments to split and compare:

```
public static void main(String[] args) {
    JaroWinklerDistance jaroWinkler = JaroWinklerDistance.JARO_WINKLER_DISTANCE;
    for (String s : args) {
        String[] pair = s.split("\\|");
        String s1 = pair[0];
        String s2 = pair[1];
        System.out.printf("%18s  %18s  %5.3f  %5.3f\n",
                          s1, s2,
                          jaroWinkler.distance(s1,s2),
                          jaroWinkler.proximity(s1,s2));
    }
}
```

**Figure 4.2** Jaro-Winkler code walk through (alias-i.com)

The algorithm compares two strings and returns a distance and proximity. The distance is the difference between two strings and the higher its value is, then more the possibility of two strings being irrelevant. The proximity is a value between 0.0 and 1.0. The higher the value returned, the higher the possibility of two strings being the same exists. A string is considered to be a sequence of characters; therefore spaces between words are blank characters and will be checked. Using a loop, the code used in the system checks only the proximity between the string submitted by the user with every question string in the database.

### 4.2.3 Combination of algorithms

In order to achieve the best possible outcome we decided to combine the two algorithms. Therefore to the database of the system has been added questions with both answers and keywords and questions with answers but no keywords. For example the keyword matching algorithm will search the keywords and the string similarity algorithm will search all pre-stored questions with or with out keywords. In order to improve this similar questions have been added that have the same answer. Basically if the keyword matching fails then we try to think the way a simple user would think, asking a question. Consider the following example in order to understand better how the combination works.

| Question | Answer | Keyword 1 | Keyword 2 |
|---|---|---|---|
| Have you received my application pack | You have to contact the admissions office for this information | Received | |
| Have you received my references | You have to contact the admissions office for this information | Received | |
| Have you received my pack | You have to contact the admissions office for this information | | |

**Table 4.3** Combination of keyword matching and string similarity

The user then submits the question: "Have you received the pack sent?"

Then the keyword matching algorithm will run first in order to identify keywords. It will identify one keyword in the first information line, one in the second and none in the third and so a deadlock occurs. In this case the string similarity algorithm will compare the pre-stored questions of the systems with the user's question and it will find that the best match is with the third entry in the table, even though there is no keyword in the database. It will then retrieve the answer of the third one. We can see here that the questions are very similar and have the same answer.

### 4.2.4 Spell checker

The system is supported by the jazzy spell checker [20]. The checker will check the sentence before it is submitted to the server. If there is an error then the user will be informed about it and about possible corrections. If it is error free then it will be submitted.

### 4.2.5 Natural language parser

The Stanford natural language parser [18] will be used to check is a sentence is valid according to specified rules. If the sentence is not valid then the user is informed and he will have to make the appropriate changes. In the case of our system a sentence is required to have a noun and a verb to be valid. The example below represents how the Stanford parser will analyse a sentence.

The user enters the question: "How can I apply?"

The parser will parse the sentence and produce a tree as follows:

```
(ROOT
  (SBARQ
    (WHADVP (WRB How))
    (SQ (MD can)
      (NP (PRP I))
      (VP (VB apply)))
    (. ?)))
```

Then the system will check that there is a noun and a verb in the tree. In the case that there is, the question is submitted to the system. In case there is not both a noun and a verb the user is informed and the user's question is not answered. The contribution of the parser is to stop the user from submitting syntactically incorrect questions. The examples below represents syntactically incorrect questions submitted by users, followed by the chatbot reply.

User: "Yes and yes not yes"
Chatbot: "My name is John."

User: "Tell me about Ph.D."
Chatbot: "Hello there."

### 4.2.6 Google search

In the case that a user is not satisfied by a system reply then he will be provided with a link from the admissions website. The link will be retrieved using customized Google search, according to the input submitted.

## 4.3 Business logic

The business logic of the system is separated in a clear way both from the database and the presentation layer. The following class diagrams describe the business logic of the system. Figure 4.2 des7cribes the implementation diagram and figure 4.3 the implemented interface.

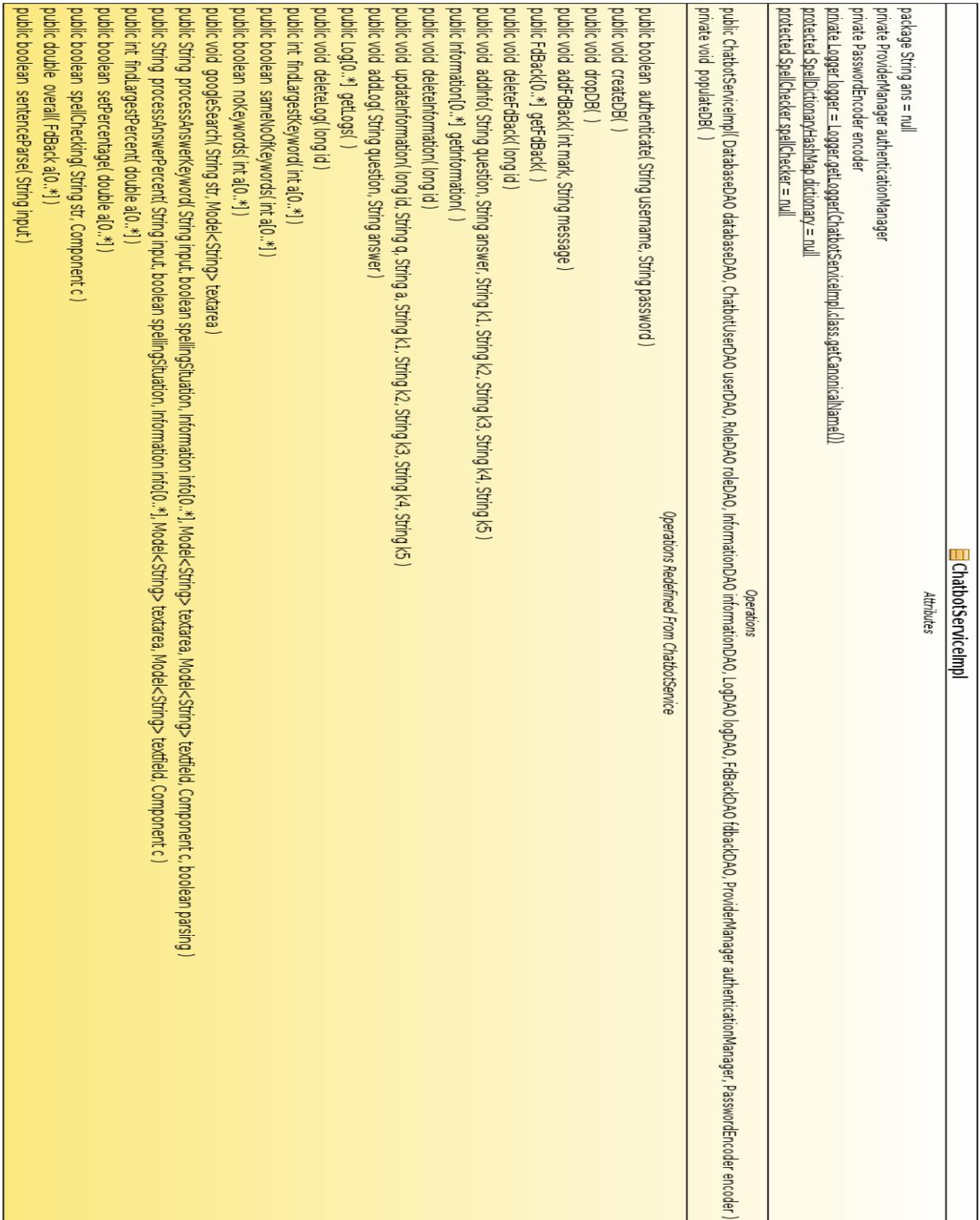

**Figure 4.3** Business Logic Implementation

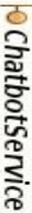

**<<interface>>**
**ChatbotService**

| Attributes |
| --- |

| Operations |
| --- |

public boolean authenticate( String username, String password )

public void dropDB( )

public void createDB( )

public void addFdBack( int mark, String message )

public FdBack[0..*] getFdBack( )

public void deleteFdBack( long id )

public void addInfo( String question, String answer, String k1, String k2, String k3, String k4, String k5 )

public Information[0..*] getInformation( )

public void deleteInformation( long id )

public void update( Information[ long id, String q, String a, String k1, String k2, String k3, String k4, String k5 )

public void addLog( String question, String answer )

public Log[0..*] getLogs( )

public void deleteLog( long id )

public int findLargestKeyword( int a[0..*] )

public boolean sameNoOfKeywords( int a[0..*] )

public boolean noKeywords( int a[0..*] )

public void googleSearch( String str, Model<String> textarea )

public String processAnswer( Keyword, String input, boolean spellingSituation, Information info[0..*], Model<String> textarea, Model<String> textfield, Component c, boolean parsing ) public String processAnswerPercent( String input, boolean spellingSituation, Information info[0..*], Model<String> textarea, Model<String> textfield, Component c )

public int findLargestPercent( double a[0..*] )

public boolean setPercentage( double a[0..*] )

public boolean spellChecking( String str, Component c )

public double overall( FdBack a[0..*] )

public boolean sentenceParse( String input )

**Figure 4.4** Business Logic Interface

## 4.4 Web Interface

The business logic of the system is separated in a clear way both from the database and the presentation layer. The following class diagrams describe the web interface of the system.

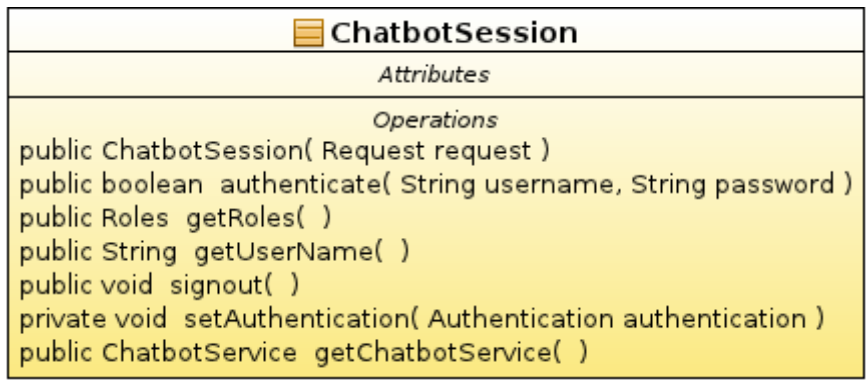

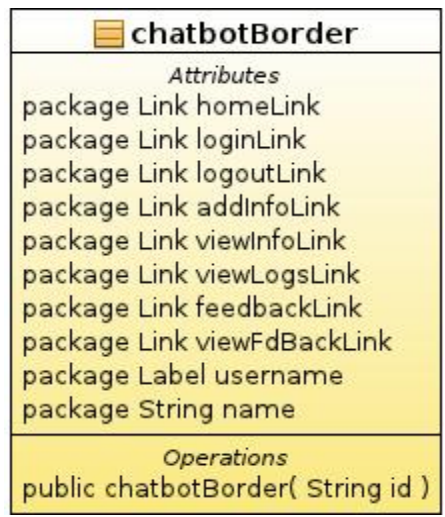

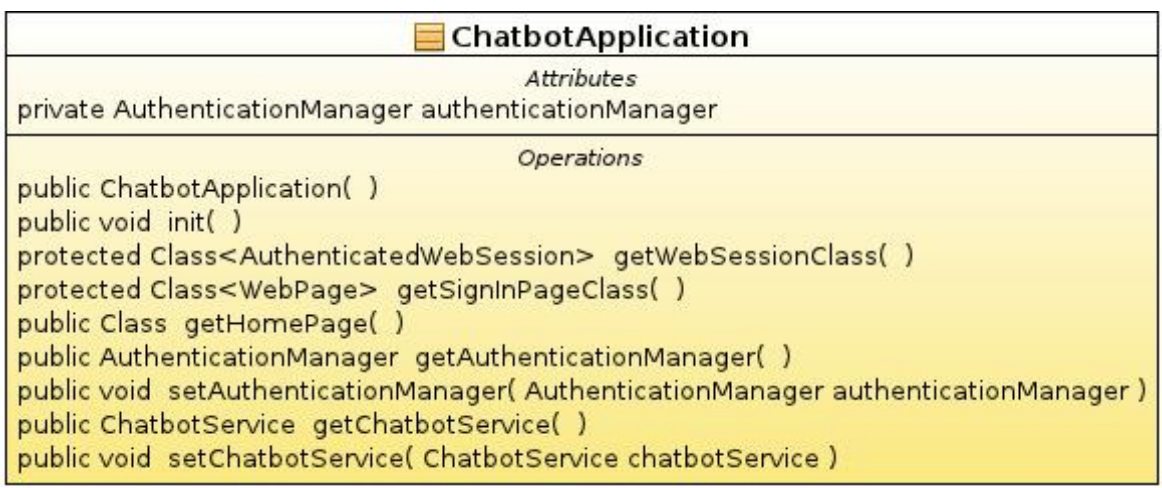

**Figure 4.5** Web interface

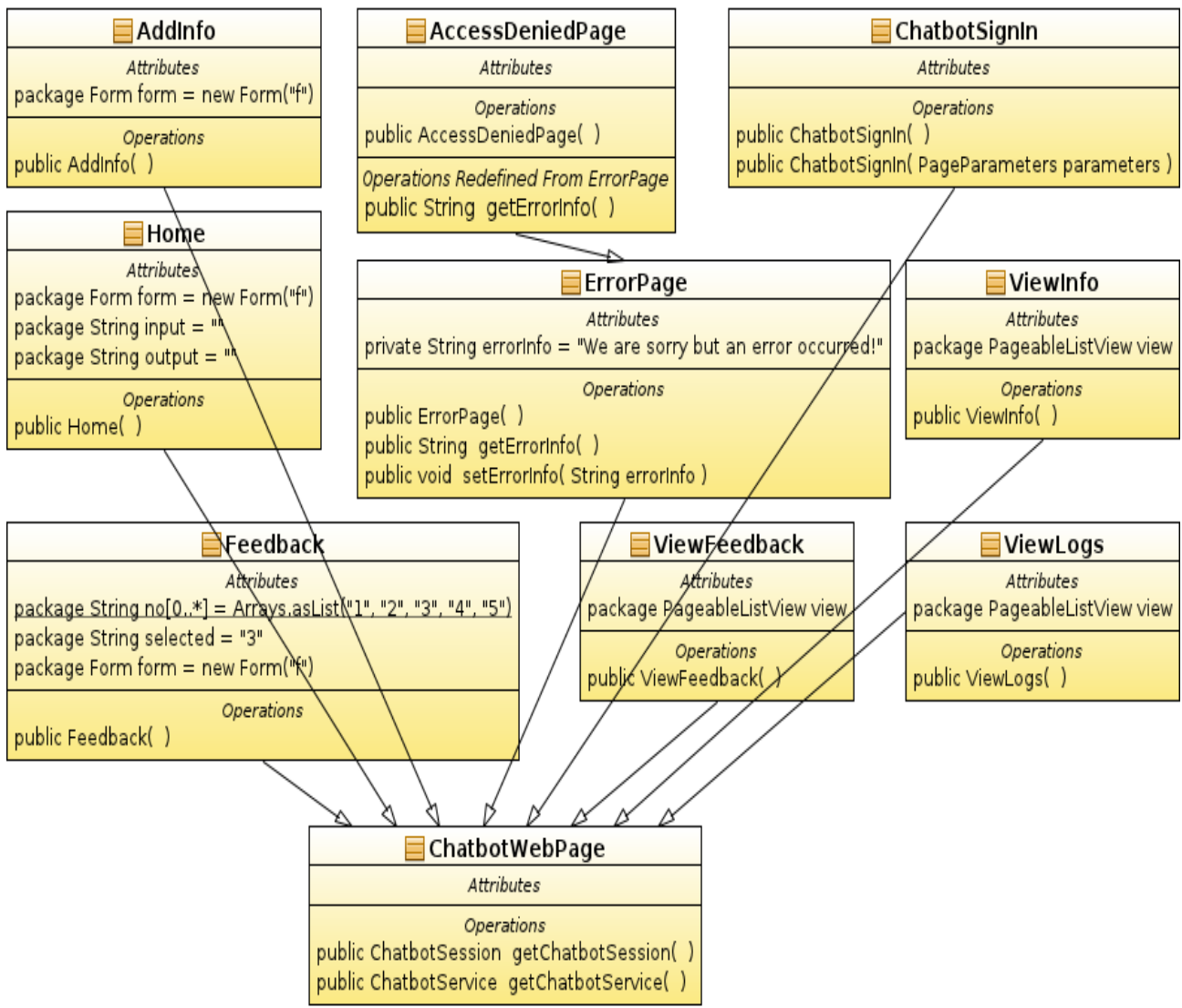

**Figure 4.6** Web interface

# Chapter 5

# Implementation

This chapter gives a description of the system implementation. User and administrator interfaces are described along with an implementation description of each component.

## 5.1 Overview

The system includes a user interface with the following options: Engage in a conversation, produce a customised link and leave feedback. The administration interface provides the options to add questions, answers and keywords. View, update and delete questions, answers and keywords. View and delete logs. View and delete feedback. The three tier architecture has been selected and described in section 2.4. This particular architecture gives the ability to develop a web application in a systematic way, where the presentation layer is separated from the application layer and the data layer. This makes sure that when a change occurs in one of the tiers the other two are not affected.

## 5.2 Database layer implementation

The first layer of the three tier architecture is the data layer, which in this case is a database. This can be achieved by using and configuring Hibernate and Spring. First a number of Java persistent objects and hibernate mapping files need to be created, followed by their configuration in the spring container, regarding the creation of the database, the connection and the associated spring data access objects.

### 5.2.1 Hibernate mapping

Mapping files have been created and used by hibernate to persist the objects to the database. Each file must be associated with one Java class. The database has five tables, each represented by a Java persistent object and its XML mapping. The code below represents a Java persistent object and its associating mapping XML file, used for the creation of the feedback table.

```java
public class FdBack implements Serializable
{
    private long id;
    private int mark;
    private String message;

    public FdBack(int mark,String message)
    {
        this.mark=mark;
        this.message=message;
    }

    protected FdBack()
    {
    }

    public long getId() {
        return id;
    }

    public void setId(long id) {
        this.id = id;
    }
    ...
```

**Figure 5.1** Java persistent object

```xml
<?xml version="1.0"?><!DOCTYPE hibernate-mapping PUBLIC
        "-//Hibernate/Hibernate Mapping DTD//EN"
        "http://hibernate.sourceforge.net/hibernate-mapping-3.0.dtd">
<hibernate-mapping package="chatbot.model">
    <class name="FdBack" table="fdback">
        <id name="id" column="id">
            <generator class="sequence"/>
        </id>
        <property name="mark" column="mark" not-null="true"/>
        <property name="message" column="message" />
    </class>
</hibernate-mapping>
```

**Figure 5.2** XML mapping

### 5.2.2 Hibernate configuration

To create the database the spring container needs to be aware of the mapping files. This is achieved in the applicationcontext.xml in a bean called sessionFactory. The figure below represents the configuration.

```xml
<bean id="sessionFactory" class="org.springframework.orm.hibernate3.LocalSessionFactoryBean">
    <property name="dataSource" ref="dataSource"/>
    <property name="hibernateProperties">
        <props>
            <prop key="hibernate.dialect">org.hibernate.dialect.PostgreSQLDialect</prop>
            <prop key="hibernate.show_sql">true</prop>
        </props>
    </property>
    <property name="mappingResources">
        <list>
            <value>chatbot/model/ChatbotUser.hbm.xml</value>
            <value>chatbot/model/Role.hbm.xml</value>
            <value>chatbot/model/Information.hbm.xml</value>
            <value>chatbot/model/Log.hbm.xml</value>
            <value>chatbot/model/FdBack.hbm.xml</value>
        </list>
    </property>
</bean>
```

**Figure 5.3** Hibernate mapping configuration

In the configuration the database is specified as PostgreSQL and the mapping resources property is used to declare the mapping files.

In the applicationcontext.xml the connection properties of the database is also included. The bean with id dataSource is used to configure the connection. The properties include a driver name, a url, a user name, a password and other connection related details. The figure below represents the connection properties.

```xml
<bean id="dataSource" class="org.apache.commons.dbcp.BasicDataSource" destroy-method="close">
    <property name="driverClassName" value="org.postgresql.Driver"/>
    <property name="url" value="jdbc:postgresql://localhost:5432/postgres"/>
    <property name="username" value="postgres"/>
    <property name="password" value="pilidis4"/>
    <property name="maxActive" value="4"/>
    <property name="maxIdle" value="4"/>
    <property name="minIdle" value="2"/>
    <property name="maxWait" value="-1"/>
    <property name="initialSize" value="2"/>
    <property name="poolPreparedStatements" value="true"/>
    <property name="maxOpenPreparedStatements" value="100"/>
    <property name="testOnBorrow" value="true"/>
    <property name="testOnReturn" value="false"/>
    <property name="testWhileIdle" value="false"/>
    <property name="validationQuery" value="SELECT 1+1"/>
</bean>
```

**Figure 5.4** Hibernate database configuration

### 5.2.3 Data access objects

A data access object is comprised from an interface and an implementation of that particular interface. Each DAO file is associated with each hibernate mapping file and its associated persistent object. An example of the feedback data access object interface and its implementation is described below.

```java
public interface FdBackDAO
{
    public FdBack loadFdBackById(long id);

    public List<FdBack> getFdBack();

    public void save(FdBack fdback);

    public FdBack delete(FdBack fdback);
}
```

**Figure 5.5** Feedback data access object interface

```java
public class FdBackDAOImpl extends HibernateDaoSupport implements FdBackDAO {

    public FdBack loadFdBackById(long id) {
        return (FdBack) getHibernateTemplate().load(FdBack.class, id);
    }

    @SuppressWarnings({"unchecked"})
    public List<FdBack> getFdBack() {
        return (List<FdBack>) getHibernateTemplate().find("from FdBack");
    }

    public void save(FdBack fdback) {
        getHibernateTemplate().save(fdback);
    }

    public FdBack delete(FdBack fdback) {
        getHibernateTemplate().delete(fdback);
        return fdback;
    }

}
```

**Figure 5.6** Feedback data access object implementation

## 5.3 Business logic layer implementation

The business or application logic of a system is considered to be one of the most important aspects. This section aims to give a clear understanding of the logic.

### 5.3.1 Business analysis

The business logic of the system is made up using two classes, chatbotService which is an interface and chatbotServiceImplementation which is the implementation of the interface. Every functional part of the system is implemented using an appropriate method. Each method is called from the presentation layer when this is necessary. In addition spring security has been used to ensure that users have access only to methods that they are supposed to. The following image represents a partial part of the interface.

```java
public interface ChatbotService
{
    public boolean authenticate(String username, String password);
    public void addFdBack(int mark,String message);
    public List<FdBack> getFdBack();
    public void deleteFdBack(long id);
    public void addInfo(String question,String answer,String k1,String k2,String
    public List<Information> getInformation();
    public void deleteInformation(long id);
    public void updateInformation(long id,String q,String a,String k1,String k2,
    public void addLog(String question,String answer);
    public List<Log> getLogs();
    public void deleteLog(long id);
    public int findLargestKeyword(int[] a);
    public boolean sameNoOfKeywords(int [] a);
    public boolean noKeywords(int[] a);
    public void googleSearch(String str,Model<String>textarea);
    . . .
```

**Figure 5.7** Chatbot service interface

The following image gives a partial description of the sentence parsing implementation method as found in the chatbot service implementation.

```java
public boolean sentenceParse(String input){
boolean tf2 =false;
try {
LexicalizedParser lp = new LexicalizedParser("englishPCFG.ser.gz");
TokenizerFactory tf = PTBTokenizer.factory(false, new WordTokenFactory());
List tokens = tf.getTokenizer(new StringReader(input)).tokenize();
lp.parse(tokens);
Tree parse = lp.apply(tokens);
parse.pennPrint();
String pn = . . .
```

**Figure 5.8** Sentence parse partial implementation

The following image gives a description of the spring security used in the system. It is described in the chatbot service security bean in the applicationcontext.xml. In the properties it is specified which methods can be accessed by the admin only.

```
<bean id="chatbotServiceSecurity"
      class="org.acegisecurity.intercept.method.aopalliance.MethodSecurityInterceptor">
    <property name="authenticationManager" ref="authenticationManager"/>
    <property name="accessDecisionManager" ref="accessDecisionManager"/>
    <property name="objectDefinitionSource">
        <value>
        chatbot.service.ChatbotService.addInfo=ROLE_admin
        chatbot.service.ChatbotService.deleteInformation=ROLE_admin
        chatbot.service.ChatbotService.updateInformation=ROLE_admin
        chatbot.service.ChatbotService.getLogs=ROLE_admin
        chatbot.service.ChatbotService.deleteLog=ROLE_admin
        chatbot.service.ChatbotService.deleteFdBack=ROLE_admin
        chatbot.service.ChatbotService.getFdBack=ROLE_admin
        </value>
    </property>
</bean>
```

**Figure 5.9** Spring security

## 5.4 Presentation layer implementation

This sections aims to give an outline of how the interface was created and how it interacts with the business logic.

### 5.4.1 Presentation analysis

The presentation layer has been developed using Apache wicket and HTML, which aims to give a clear separation between the template and the business logic. This is achieved by using appropriate wicket ids in the HTML code and give actions to these components in the Java code were the id is associated. Moreover when an action needs to be provided the appropriate method is called from the chatbot service interface. In order to achieve that a bean needs to be declared in the applicationcontext.xml as shown below. The bean is called wicket application and then chatbot service is instantiated. This gives us the ability to call the chatbot service interface in the presentation and subsequently call any of the authorised methods.

```
<bean id="wicketApplication" class="chatbot.page.ChatbotApplication">
    <property name="authenticationManager" ref="authenticationManager"/>
    <property name="chatbotService" ref="chatbotService"/>
</bean>
```

**Figure 5.10** Spring Integration with Wicket

## 5.5 User Interface

The final system includes a web interface with the following options: Chat, Login, Leave feedback, add information, update information, delete information, view and delete feedback and view and delete logs. This sections shows the interface of the system.

### 5.5.1 Chatbot interface

The home page of the system where the user can engage in a conversation is shown below.

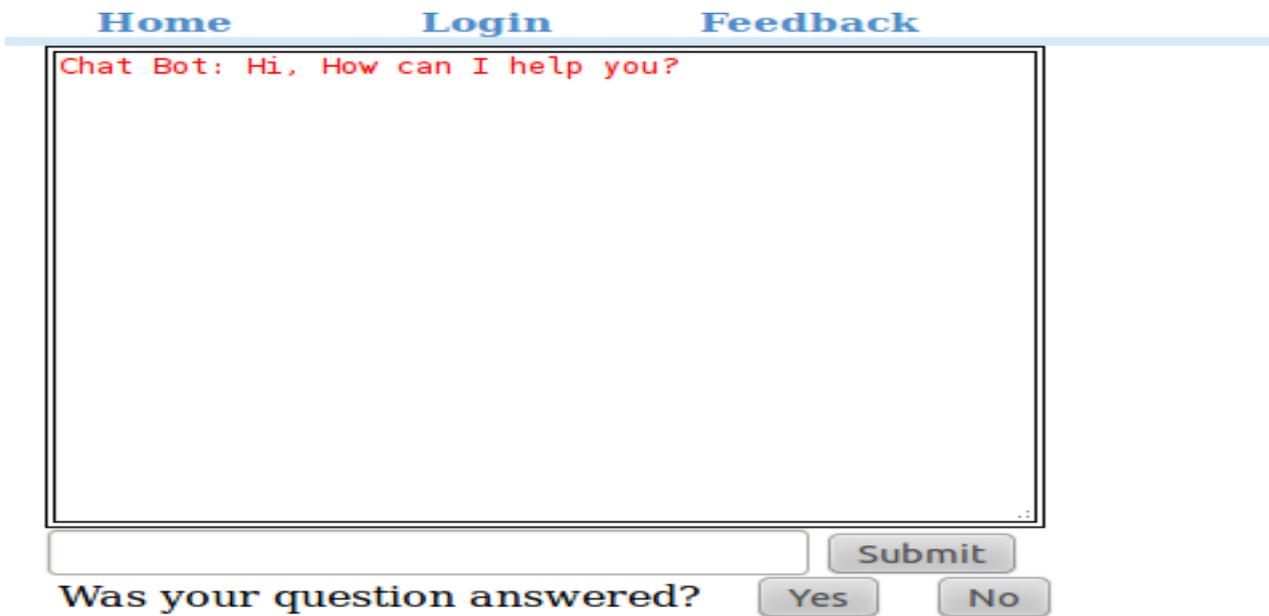

**Figure 5.11** Chatbot

### 5.5.2 Leave feedback

The feedback page of the system where the user can leave a mark and feedback message is shown below.

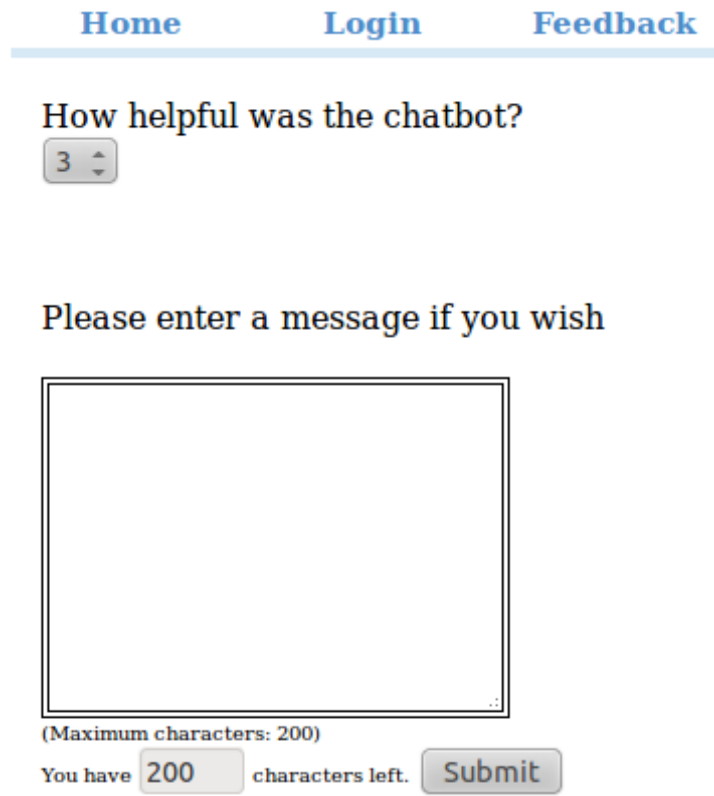

**Figure 5.12** Feedback

### 5.5.3 Add information

The add information page of the system where the administrator can add questions, answers and keywords is shown below.

**Figure 5.13** Add information

### 5.5.4 View information

The view information page of the system where the administrator can view, update and delete questions, answers and keywords is shown below.

**Figure 5.14** View information

### 5.5.5 View feedback

The view feedback page of the system where the administrator can view and delete feedback messages is shown below.

**Figure 5.15** View feedback

### 5.5.6 View logs

The view logs page of the system where the administrator can view and delete log messages is shown below.

**Figure 5.16** View logs

# Chapter 6

# Testing and Evaluation

This chapter is divided into two sub chapters. The first part describes the testing of the system and the test cases. The second part describes how the system was evaluated.

## 6.1 Testing

It is vital in a software system that it behaves as expected. To ensure that the system and its components will behave as required a number of tests was taken.

### 6.1.1 Overview

Every unit of the system was tested to make sure that the actual result of its operation was what we expected. Furthermore when every unit was integrated with other units during the development it was tested again to check its operation with other system units.

### 6.1.2 Unit and integration testing

A testing strategy known as unit and integration testing has been used to check that the system behaves as expected. The testing strategy was based on the functionality and the requirements of the system. The testing table below gives a precise description of the tests that took place.

| Test case id | Test description | Expected result | Actual result | Result |
|:---:|:---|:---|:---|:---:|
| TC001 | Ask a question | Provide answer | Answer provided | PASS |
| TC002 | Ask a question with a spelling error | Spell checker identified the error | Spell checker identified the error | PASS |
| TC003 | Enter an invalid sentence | Sentence parser identified the error | Sentence parser identified the error | PASS |
| TC004 | User satisfied by the answer | Nothing happened | Nothing happened | PASS |
| TC005 | User not satisfied by the answer | A log of the question and answer kept and a relevant link provided | A log of the question and answer kept and a relevant link provided | PASS |
| TC006 | User submits feedback mark and message | Feedback submitted | Feedback submitted | PASS |
| TC007 | Administrator logs in | Log in successful | Log in successful | PASS |
| TC008 | Administrator | Information added | Information added | PASS |
| | adds new information | successfully | successfully | |
| TC009 | Administrator views information | Information displayed | Information displayed | PASS |
| TC010 | Administrator updates information | Information updated | Information updated | PASS |
| TC011 | Administrator deletes information | Information deleted | Information deleted | PASS |
| TC012 | Administrator views logs | Logs displayed | Logs displayed | PASS |
| TC013 | Administrator deletes a log | Log deleted | Log deleted | PASS |
| TC014 | Administrator views feedback | The overall rank of the system and the feedback messages displayed | The overall rank of the system and the feedback messages displayed | PASS |
| TC015 | Administrator deletes a feedback | Feedback deleted | Feedback deleted | PASS |
| TC016 | Administrator logs out | Log out succeeded | Log out succeeded | PASS |

**Table 6.1** Unit and Integration testing

### 6.1.3 Performance testing

Every component developed passed through a performance test to ensure that the system meets certain performance requirements. Finally when each unit was developed and tested then it was integrated with the rest of the system to do a performance test. The figure below describes the stages of performance testing [19].

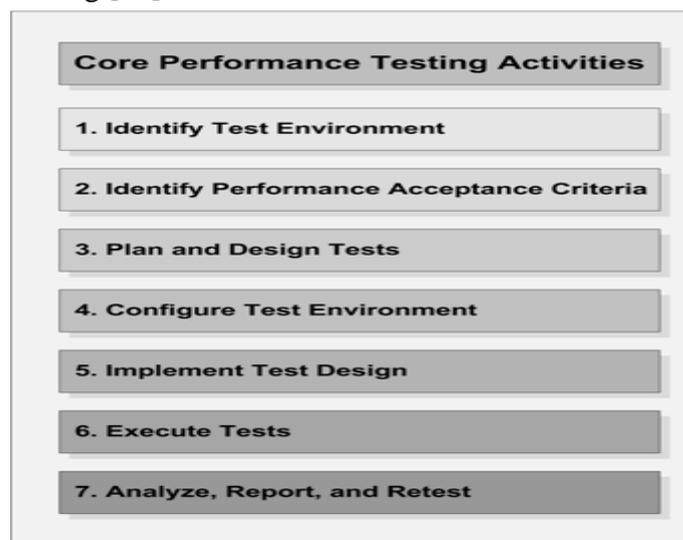

**Figure 6.1** Performance testing stages

Every component of the system went through each stage of the approached methodology. It was identified that each unit behaved differently as a single entity and differently when it was integrated with other units in respect of speed. The test environment plays a crucial role on the performance of the system. A different result occurs when the system runs on development mode and a different one when the system has been deployed. Moreover the technical specifications of the server that the final system has been deployed play their role as well.

### 6.1.4 Security testing

Security testing refers to ensuring that the system will protect its data from unauthorized access and modification and that it will continue to behave as expected. Different security concepts have been used to ensure that the system is secure. Authentication has been used to confirm the identity of someone being either a user or an administrator. Authorisation has been used to ensure that users have access to pages that they are supposed to and that the administrator has an overall control of the system. Access control has been used to achieve that. To make sure that the system is safe from attacks, including SQL injection and XSS, a number of tests took place and are shown below.

| Test case id | Test description | Expected result | Actual result | Result |
|---|---|---|---|---|
| TC001 | User knows the exact link of the view information page and tries to submit it directly | System redirects to the log in page | System redirects to the log in page | PASS |
| TC002 | User tries to submit an XSS attack script through the text area | The system skips the relevant characters and stops the attack | The system skips the relevant characters and stops the attack | PASS |
| TC003 | User tries to submit an SQL injection attack script through the text area | The system skips the relevant characters and stops the attack | The system skips the relevant characters and stops the attack | PASS |
| TC004 | The user submits random user name and password | The system rejects the log in | The system rejects the log in | PASS |
| TC005 | The user gains access to the business logic and tries to execute certain methods | The user is not authorised to do that and it is rejected | The user is not authorised to do that and it is rejected | PASS |

**Table 6.2** Security testing

## 6.2 Evaluation

Evaluation is the process where we collect and analyse information in order to find out if the software system developed is effective and efficient.

### 6.2.1 Overview

A major part of this project was to build a usable system, based on real world requirements and to evaluate it using potential students. The system managed to satisfy every requirement, including new requirements introduced after feedback received from potential students. However due to time constraints it was not possible to develop an Android part of the project.

### 6.2.2 Question categories

For the purposes of analysis, the types of questions that a user could possibly ask have been categorised into four main categories.

**Relevant question:** A relevant question refers to a valid question. The answer exists in the database and the algorithm is able to retrieve it.

**Irrelevant question:** An irrelevant question refers to a question that is not related to the admissions processes and has not been included in the database.

**No response question:** A no response question refers to a valid question, however the answer does not exist at the moment in the database.

**Poor response question:** A poor response would be that the answer exists in the database but the algorithm would fail to retrieve it and would come back with an incorrect or irrelevant reply.

Examples of each of these categories are given below.

A relevant question that a user asked: "What are the entry requirements for the computer security MSc?"
The reply was correct: "To enter the programme, you need to have at least an Upper Second Class (2.1) degree or an international equivalent in Computer Science or a closely related discipline and a solid foundation in programming."

An irrelevant question asked: "Size of university grounds in square miles?"
The reply was irrelevant: "Yes, with most of the major players such as HP,IBM and Logica. Which is the reply to the question: Are there any links with the industry?"

A no response question asked: "Where is the toilet?"
The reply was irrelevant: "The starting dates for the programme are: 2011/12: 26th September 2011, 2012/13: 24th September 2012.It is not possible to join the programme in January or at any other time after the start date."

A poor response question asked: "Is it OK to pay my fees using a visa?"
The reply was poor: "Yes, unless you are from an EU or an EEA country. If you require a visa and have accepted your offer you will receive a cas statement by email soon.
The answer for the question existed on the database; however the retrieved answer is the answer to the question: Do I need a visa to study in the UK? "

### 6.2.3 User evaluation

The system was evaluated by asking current applicants from Masters and undergraduate programmes to test it over the WWW. The messages were sent to MSc applicants with:

1. Unconditional offers (not yet accepted)
2. Unconditional offers (accepted)
3. Conditional offers (not yet accepted)
4. Conditional offers (accepted)
5. Applications that still required references and/or transcripts

Messages were sent to two sets of MSc applicants:

1. Feasibility study: 108 emails were sent to 74 applicants with surnames beginning with S or T.
2. Main evaluation: 770 emails were sent to 537 applicants with surnames beginning with any other letter.

Messages were sent in two sets of undergraduate applicants:

1. Feasibility study: Emails were sent to 18 applicants with surnames beginning with S or T and who had offers.
2. Main evaluation: Emails were sent to 75 applicants with surnames beginning with any other letter and who had unconditional offers.

Approximately eighty MSc students used the system and left feedback either using the form on the website or by email. Unfortunately only two undergraduate students used the system. Both positive and negative messages were received from the users. Most of the messages were about the knowledge of the chatbot and not being able to answer a number of questions, because an answer did not exist in the database. Moreover a small number of questions could not be answered by the chatbot because the algorithm could not retrieve the answer. Other replies were referring to the fact that the chatbot can not answer specific questions after acceptance and that the user interface is not of a satisfying level.

A selection of the feedback messages can be found in appendix D.

### 6.2.4 Survey

This section aims to analyse the results collected from the logs and the feedback from the period that the system was online. There are two result sets collected from both undergraduate and postgraduate students. The initial result, that is based on the results collected from an initial 7 day period and from applicants with surnames beginning with S or T, the results of which were used to improve the system and upload it a second time to collect feedback from applicants with surnames beginning with any other letter. The final system was hosted for 15 days.

Feedback was collected by the feedback page of the system, where the user could give a 1-5 scale score and leave comments if desired, and by emails sent by the applicants.

The tables and pie charts in the initial results and final results sections below, show the user response when they felt the system had failed. They do not represent the performance of the whole system.

**6.2.4.1 Initial results**

Through the feedback page of the system four messages were received, with an overall score of 2.75. Furthermore another four messages were received by email that only contained suggestions about the system.

An assumption has been made through the feedback received and by the number of logs that approximately 10 people have used the system. The table below represents the categories and the number of questions not answered by the system for each category.

| Irrelevant | No response | Poor response |
|:---:|:---:|:---:|
| 12 | 12 | 8 |

**Table 6.3** Initial logs

The pie chart below represents the percentages from the number of logs from the database of the system, during its first use.

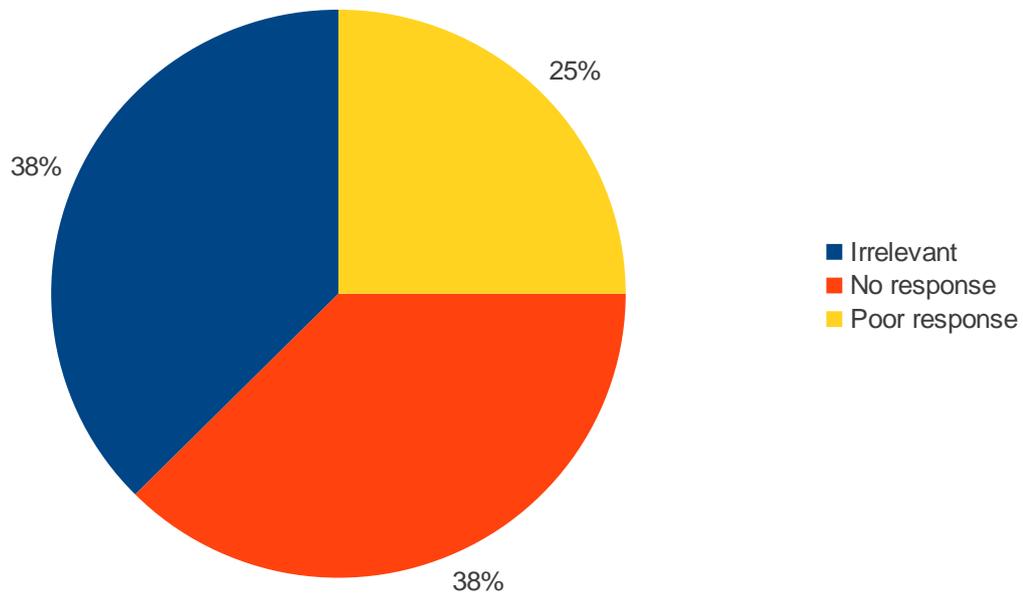

**Figure 6.2** Initial log results

After the initial use and by careful consideration of the feedback messages a number of changes were made and a final system was introduced. The main changes after the initial evaluation of the system were:

**Natural language parsing:** It was decided that the use of a sentence parser was necessary for the system, the parser would analyse the structure of a sentence and identify that a noun and a verb exists, thus avoiding submitting syntactically incorrect questions to the system. After careful consideration the Stanford parser was chosen.

**Database enrichment:** After the initial use it was clear that the information in the database was not of a satisfactory level and more data had to be added to it. A high number of relevant admission questions found in the logs and a response did not exist in the database.

### 6.2.4.2 Final results

Through the feedback page of the system thirteen messages were received, with an overall score of 2.53. Furthermore another twenty two messages were received by email that only contained suggestions about the system.

An assumption has been made through the feedback received and by the number of logs that approximately 50 people have used the system. The table below represents the categories and the number of questions not answered by the system for each category.

There were three cases for which a relevant answer was provided, however the user was not satisfied.

| Relevant | Irrelevant | No response | Poor response |
|----------|-----------|-------------|---------------|
| 3 | 62 | 61 | 31 |

**Table 6.4** Final logs

The pie chart below represents the percentages from the number of logs from the database of the system, during the final use.

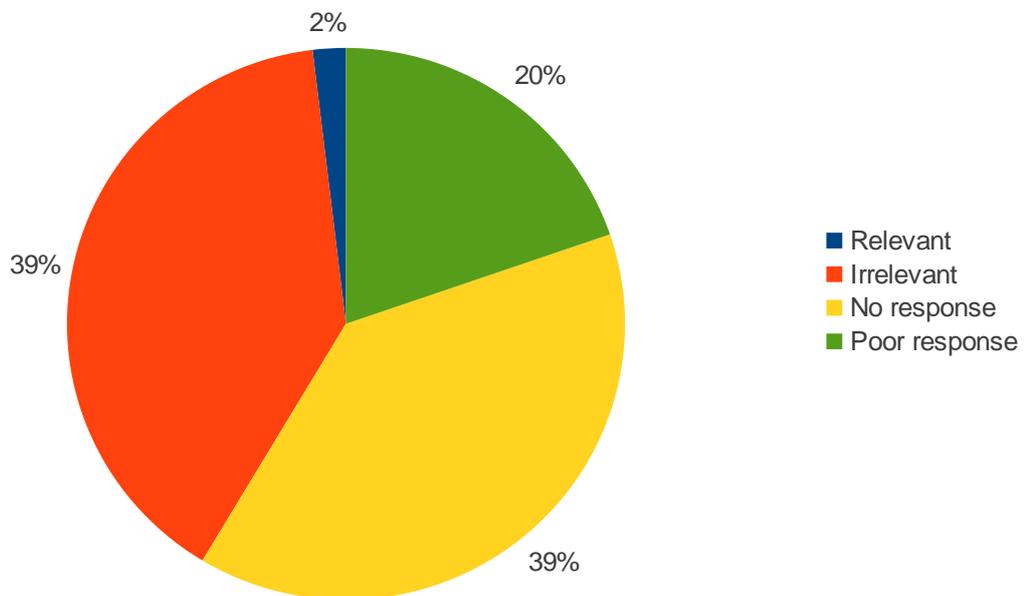

**Figure 6.3** Final log results

The suggestions from the final feedback have been taken into consideration and the most important ones that will make a better system. These include:

**User Interface:** The user interface should be more interactive and include more details.

**Database enrichment:** After the initial use it was clear that the information in the database was not of a satisfactory level and more data had to be added to it. A high number of relevant admission questions and answers were added, however there is still much to be added.

**Precise answers:** Answers must be more specific and include more details.

**Information colour:** The colour of the question and the answer must be different in the chatbot text area.

More details regarding future work can be found in section 7.3.

# Chapter 7

# Summary

This chapter refers to the summary and conclusion. Firstly it discusses the positive points of the system, followed by its limitations, future work and conclusions.

## 7.1 Achievements

A number of positive points were identified. A working system was developed using Java and a number of third party libraries. The system was uploaded on two occasions on the schools web server. The first time was used by a small number of potential students, which left feedback and some modifications took place. The second time a larger number of potential students used the system and feedback was taken. An interesting part was the logs kept by the system that revealed how people interact with a chatbot.

The fact that the system used a spell checker is considered to be a positive aspect, since the possibility of retrieving the wrong answer due to an error is reduced. Moreover a sentence parser was used to inform the user if a sentence is valid or not, before submitting. A customized google search functionality that provides a link relevant to the question, if the user was not happy by the answer and a log of the question and the answer is kept at this stage. Moreover the user has the option to submit a feedback message. Another positive thing is that the web application is dynamic and has an administrator interface. An administrator has the option to log in and make appropriate changes in a number of ways. The administrator has the option to add information to the database, such as questions, answers and keywords. Information already in the system can be viewed, updated and deleted. Also the option to view feedback messages and delete them is available. The option to view and delete logs is also available. Another positive aspect is that the system is secure from unauthorized access and from various attacks.

The results gathered from the users helped to identify the limitations of the system and make changes for a number of them, when time permitted. The evaluation results in combination with the testing strategy used helped to make the system as effective as possible.

## 7.2 Limitations

A limitation on the algorithm of the system was found after evaluation and testing. The fact is that the possibility of not retrieving the correct answer exists. A small number of the logs were about this problem. Moreover potential students want to know about a different number of things and not everything is covered in the database. A problem is that the system has no memory and when a user asks the same question for a number of times the answer remains the same. Anaphora resolution is not present, which would be useful in some situation when a user refers to a previous question. However anaphora resolution would require a chatbot memory to be available. A limitation in the spell checking is that the system will suggest a word for a spelling error; however the user will have to change that manually and submit the question again. Although there might be a number of questions that do not require a verb the sentence parser will reject every question that does not contain a noun and a verb and the user will have to restructure the sentence and resubmit it. Regarding the Google customised search the links provided are not hyper linked. The fact that the system is usable was an achievement however the user interface should be improved and be more interactive.

## 7.3 Future work

A number of different algorithms should be implemented, tested and evaluated. Template matching should be a high priority algorithm. Furthermore algorithms could be used in combination and specify a percentage for each. A possible solution would be to upload to the web server each implementation separately and collect feedback from users. This will make easier to identify the limitations of each algorithm and come with the best solution that can also be a combination of two or more algorithms. There is also a different number of factors that should be taken into consideration except the retrieval, such as speed and compatibility. Also more advanced natural language processing techniques if used will help to identify a possible answer. Anaphora resolution will help to identify when a user refers to a previous question or answer, which requires the user of a chatbot memory. A number of techniques such as statistical parsing and analysis of the sentence structure will help to identify the meaning of a sentence and provide better answers. Knowledge based information retrieval will help to find an answer to a particular question in the frequently asked questions of the admissions website in the case that the algorithm will not find a particular answer. Also the customised google search helps in that direction with the disadvantage that in a number of cases the link provided might not be the one required.

The logs that the system keeps is an important aspect of the system, which will help the administrator to find user questions not answered because the particular query and answer does not exist in the database. It is therefore important to maintain a database updated continuously that aims to cover most of the users' queries, both undergraduate and postgraduate. It would be helpful to create a statistics page and database table where the number of correctly answered and not correctly answered questions will be stored. It is very helpful to find how efficient the system is.

There is also a number of small changes that could improve the system in general and do not depend on the natural language processing used. The use of a customised search using an actual search text area will be really useful for the user. The interface of the system should be improved and become more interactive and additional information such as links to the university, the school and to the admissions FAQs should be added.

Regarding the spell checker it would be more useful when an error is found to provide a hyper linked version of the sentence with the error fixed so the user can have a better and faster experience. Also the case when the user asks more than one question at the same time should be considered and the possibilities of using certain rules and provide separate answers or informing the user to submit only one question at time. Moreover a more comprehensive dictionary should be used to cover more varieties of English and not U.K only.

The option to make a chatbot to learn from users exists if certain rules are used and can be divided into two categories. Learn and use only for the current session and if the user asks the same question or a similar then provide the answer. Moreover this information can be saved to the database and be used to reply to future users. This can be done by just saving or by saving and then comparing to other saved information and decides if it is correct or not. However this is not recommended for an admissions system where every answer must be precise.

Also a high number of overseas students wish to apply to the university and at the time of the application the level of English that they process might not be sufficient for a conversation. There are two different aspects regarding the use of foreign languages. The user might not be able to provide a syntactically correct sentence and the parser will reject it. However the keyword matching will identify the important words if the sentence goes through the sentence parser to the keyword matching.

## 7.4 Conclusions

The main objectives of the project were to develop an algorithm that will be used to identify answers related to user submitted questions. To develop a database were all the related data will be stored and to develop a web interface. The web interface developed had two parts, one for simple users and one for the administrator.

The following objectives were subsequently met. A background research took place, which included an overview of the conversation procedure and any relevant chatbots available. A database was developed, which stores information about questions, answers, keywords, logs and feedback messages. A usable system was designed, developed and deployed to the web server on two occasions. An evaluation took place from data collected by potential students of the University. Also after received feedback from the first deployment, extra requirements were introduced and implemented.

# References

1. Cawsey, Alison, 1998. The Essence of Artificial Intelligence. London: Pearson and Prentice Hall

2. Negnevitsky, Michael, 2005. Artificial Intelligence: A guide to intelligent systems, 2nd ed. London: Pearson

3. Tong Ka Iok, Kent, 2010. Enjoying Web Development with Wicket. 2nd ed. London: lulu.com

4. Marnie L Hutcheson 2003. Software testing fundamentals. London: John Wiley and sons

5. Michael L. Mauldin, Carnegie Mellon University Centre for Machine Translation Chatterbots, Tinymuds, and the Turing Test Entering the Loebner Prize Competition. Available at: http://www.aaai.org/Papers/AAAI/1994/AAAI94-003.pdf
[Accessed on 13 June 2011]

6. Artificial Intelligence Foundation. A.L.I.C.E. Available at: http://alice.pandorabots.com/
[Accessed 15 June 2011]

7. Eliza Chat Bot. Available at: http://nlp-addiction.com/eliza/
[Accessed 15 June 2011]

8. University of Birmingham, School of Computer Science admissions. Available at: http://www.cs.bham.ac.uk/admissions/ [Accessed 15 June 2011]

9. School of Computer Science University of Birmingham. Internet Computing Workshop part 2 by Behzad Bordbar. Available at: http://www.cs.bham.ac.uk/~bxb/Teaching.html#iss
[Accessed 20 June 2011]

10. W3Schools HTML tutorial. Available at: http://www.w3schools.com/html/ [Accessed 20 June 2011]

11. The waterfall model. Available at: http://www.selectbs.com/analysis-and-design/what-is-the-waterfall-model [Accessed 20 June 2011]

12. Content with style. A CSS framework by Mike Stenhouse. Available at: http://www.contentwithstyle.co.uk/content/a-css-framework/ [Accessed 20 June 2011]

13. Apache Wicket component reference. Available at: http://wicketstuff.org/wicket14/compref/
[Accessed 21 June 2011]

14. Google Java spelling api. Available at: http://code.google.com/p/google-api-spelling-java/
[Accessed 25 June 2011]

15. Google search Java sdk. Available at: http://code.google.com/p/google-search-java-sdk/
[Accessed 1 July 2011]

16. Spring Security and Wicket-auth-roles. Available at: https://cwiki.apache.org/WICKET/spring-security-and-wicket-auth-roles.html
[Accessed 7 July 2011]

17. Lingpipe Java NLP Library. Available at: www.*alias-i.com/lingpipe/*
[Accessed 10 July 2011]

18. The Stanford Library. Available at: http://nlp.stanford.edu/
[Accessed 15 August 2011]

19. Performance testing for Web Applications. Available at: http://msdn.microsoft.com/en-us/library/bb924375.aspx
[Accessed 25 August 2011]

20. The Jazzy spell checker Library. Available at: http://jazzy.sourceforge.net/
[Accessed 25 August 2011]

21. Bayan Abu Shawar, Eric Atwell, Chatbots: Are they really useful? Penn state University. Available at: http://citeseerx.ist.psu.edu/viewdoc/download doi=10.1.1.106.1099&rep=rep1&type=pdf
[Accessed on 26 August 2011]

22. Ehab El-agizy, Moustafa Zamzam, AI Chatbots developing. Available at:

http://www.infradrive.com/downloads/articles/Article1.pdf
[Accessed on 27 August 2011]

# Appendix A Project proposal

### 1. Introduction
This document covers an introduction to the project including the context, a description of aims and objectives and a sound plan for completion.

### 2. Context
Experience has shown that applicants are more likely to choose to come to a department where they believe that they are treated as a valued individual. The opportunity to communicate with on a one-to-one basis is highly valued. However with many hundreds of applications each year, one-to-one conversations are not feasible in most cases.

Although the admissions process works properly as it is, it is very difficult and time consuming to contact a member of staff of the university. However, the problem would be partially solved if the applicant could talk to a convincing chatbot, able to respond to their concerns with information about admissions, booking accommodation, paying fees in instalments and what pre-sessional courses are on offer.

By the end of the development, the chatbot application will be able to communicate with a user in a way similar to the following:

Chat Bot: Hello how can I help you?
User: What is the minimum score of IELTS required for entry into MSc Computer Science?
Chat Bot: The minimum requirement is a total score of 6.0 with no less that 5.5 in any section.
User: What other test is accepted by the department?
Chat Bot: We also accept TOEFL and the certificate of Proficiency of the University of Cambridge.

### 3. Aims and Objectives
The aim of this project is to contribute to the solution of the problem of direct communication between applicants and the university.

The main objectives of the project are as follows:

**Information resource:** To develop an information resource were all the relevant information about the questions and the answers will be stored.

**Algorithm:** To develop an algorithm that will be able to identify meaningful keywords in the users' input and provide a suitable answer.

**Interface:** To develop a web interface which aims to give the ability to potential students and their families to submit questions in a chatbot and get convincing replies. The web system can be hosted on any computer that supports an operating system on which the Java Virtual machine will be able to load. Furthermore the Apache Wicket and the Google spelling libraries should be available. The porting of the application to the Android platform will require the use of the Android simulator and the Google spelling library.

## 4. Project Plan

| | Activities |
|---|---|
| Week 1<br>13-19 June | 1. Research about chatbot systems in general<br>2. Use ELIZA and A.L.I.C.E<br>3. Research NLP in Java and try Open Source spell checkers |
| Week 2<br>20-26 June | 1. Start developing the web interface<br>2. Research how chatbots react to sub questions |
| Week 3<br>27 June – 3 July | 1. Begin developing the algorithm and create a basic version<br>2. Connect the algorithm with the web interface<br>3. Test the basic system and collect feedback for improvement |
| Week 4<br>4-10 July | 1. Continue developing the algorithm<br>2. Finalize the web interface |
| Week 5<br>11-17 July | 1. A complete version of the web interface is now available<br>2. Test the algorithm to see how it behaves when the user jumps from a topic to another in a way that is difficult to understand |
| Week 6<br>18-24 July | 1. Include an open source spell checker to the application<br>2. Test the spell checker<br>3. Continue developing the algorithm |
| Week 7<br>25-31 July | 1. A complete version of the algorithm is now available and works with the spell checker and the web interface properly<br>2. Continue testing on the final version to identify possible errors |
| Week 8<br>1-7 August | 1. Upload the system to the web server<br>2. Test the system in the web server context to see how it behaves |
| Week 9<br>8-14 August | 1. Port the system to the android platform<br>2. Test the android version of the system |
| Week 10<br>15-21 August | 1. Continue testing both systems<br>2. Final enhancements of the system |
| Week 11  22-28 August | 1. Prepare the presentation |

# Appendix B How to run the system

The system has been developed using Netbeans and it is recommended to open the project directory using Netbeans. The project directory can be found in the CD and it is called "Chatbot".

The files below can be found under the "Chatbot" directory in the "src" directory.

Under the default package the ApplicationContext.xml file can be found. It is the Spring configuration file, which uses the hibernate mapping files to create the database, set up the beans, manage transactions, authentication and security. In that file there exists a bean with the name datasource. We have to change a number of properties to comply with our system. The url property needs to be changed to map to our database, the username need to be changed with the username of our database and the password needs to be changed with the password of our database.

```
<bean id="dataSource" class="org.apache.commons.dbcp.BasicDataSource" destroy-
method="close">
    <property name="driverClassName" value="org.postgresql.Driver"/>
    <property name="url" value="jdbc:postgresql://localhost:5432/postgres"/>
    <property name="username" value="postgres"/>
    <property name="password" value="pilidis4"/>
    <property name="maxActive" value="4"/>
    <property name="maxIdle" value="4"/>
    <property name="minIdle" value="2"/>
    <property name="maxWait" value="-1"/>
    <property name="initialSize" value="2"/>
    <property name="poolPreparedStatements" value="true"/>
    <property name="maxOpenPreparedStatements" value="100"/>
    <property name="testOnBorrow" value="true"/>
    <property name="testOnReturn" value="false"/>
    <property name="testWhileIdle" value="false"/>
    <property name="validationQuery" value="SELECT 1+1"/>
</bean>
```

After this is done we have to run the Main file under the /chatbot/ sub directory. The it will pop up a message as shown in the figure below.

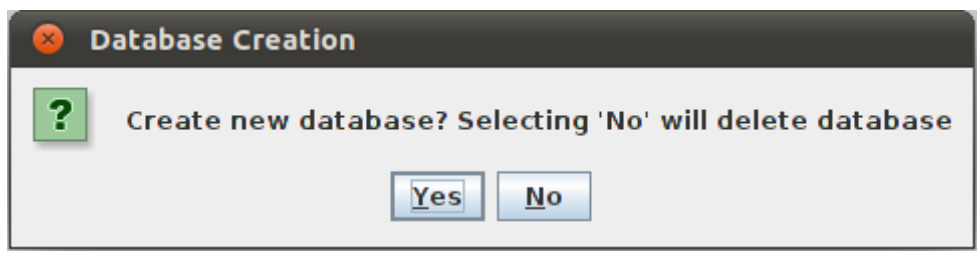

If the Yes button is clicked then the tables are created and a number of data fill these tables. If the no button is clicked the database tables are deleted if exist. The no button should be clicked if the database has been created by clicking yes and we have made changes so it is necessary to delete it and re create it.

The next step is to Clean and build the project and run it.

The other directories of the project are as follows:
The /chatbot.dao/ subdirectory stores the interfaces of the data access objects that spring uses to access the database.

The /Chatbot.dao.hibernate/ subdirectory stores the implementation of the previous interfaces.

The /Chatbot.model/ subdirectory stores the persistent classes and the hibernate mapping files.

The /Chatbot.page/ subdirectory stores the Java and HTML files for the presentation layer used by wicket.

The /Chatbot.service/ stores the chatbot service interface and the chatbot service implementation that contain the business logic of the project.

**Furthermore on the CD the following files can be found:**

The eng_com.dic is the dictionary used for the spell checker. In the chatbot service implementation file in the spell Checking method the location of this file needs to be specified.

The dissertation.odt is the report of the project, written using open office.

The dissertation.doc is a Microsoft word version generated from open office.

The dissertation.pdf is the pdf version of the report.

The front page .odt is the front page of the report.

# Appendix C Example Conversation

The figures below display how an example conversation would appear if a user decided to engage a conversation with the chatbot.

Home                    Login                    Feedback

Chat Bot: Hi, How can I help you?
User: What are the entry requirements for the computer security msc?
Chat Bot:To enter the programme, you need to have at least an Upper Second Class (2.1) degree or an international equivalent in Computer Science or a closely related discipline and a solid foundation in programming.

How can i find out about scholarships?    Submit

Was your question answered?    Yes    No

Home                    Login                    Feedback

Chat Bot: Hi, How can I help you?
User: What are the entry requirements for the computer security msc?
Chat Bot:To enter the programme, you need to have at least an Upper Second Class (2.1) degree or an international equivalent in Computer Science or a closely related discipline and a solid foundation in programming.
User: How can i find out about scholarships?
Chat Bot:For PhD scholarships funded by the school every application is automatically considered. For other PhD and MSc that exist you need to apply for each seperately. Further information can be found here: http://www.cs.bham.ac.uk/admissions/

Submit

Was your question answered?    Yes    No

# Appendix D Feedback

**User 1**

I have tried the system out and found it to be quite informative, I hope this helps.

**User 2**

I don't know how helpful this would be to the project, but the questions I asked this bot are (had these when first applied (answered through website)):

General Comments:
It may help to have a follow up question concerning the course chosen when an answer would differ according to masters courses. It also would seem that having mentioned 'Masters', causes a priority above other nouns that may be more specific to the question at hand. There are answers about the actual applying of courses, but often not  answers about the specifics of taking the course itself.

Questions not answered:
Do I have to do research for the project before the summer? / When
do I have to do research for the project?
When do I have to get a supervisor for the Project?
How long do I have to choose the modules? / When do I have to choose modules?
Can I repeat modules that I have already completed in my degree?

External Applicants:
What computers can I use?
Does it cost money to print?
What computers can I use?
Chat Bot:My name is John
What happens if I send in work late?
Who can I email about questions I have about the course?
How do I ask questions about something in a module?

**User 3**

Firstly, is the GUI, in my browser ,the interface shows incomplete, as follows:

I hope this system can use for all browsers.

Secondly, I hope this system can add a flash in order to  guide the user to use.

Thirdly, I hope this system can provide a link to the main page of university and give the user a contact E-mail address.

Actually I think the main interface is too simple, as a user the first feeling is not good.

In addition, can you give me  a account to test the main function?

**User 4**

Good day. Quite an impressive project. It does answer correctly properly structured questions especially ones related to admissions, application contact information. There are a few spelling mistakes e.g. reply to Can I have a virtual tour of the computing facilities?

Chat Bot:The school is housed in a modern building with 24 hour access a day, 7 days a weak. There are dedicated laboratories for teaching,robotics,medical imaging, a library and a wireless network.

Is there any reason why the links displayed as answers are not hyper-linked?

**User 5**

So far as I can see, the answers that the chat bot has are very limited (maybe this is because the creator hopes the bot to gives general answer only). There are still a lot of rooms to improve, such as the preprocessing (do not suggest words unrelated to the context, such as "boat"), variation uses of question words (what, why, where, etc.) and the other words, part of speech tagging, etc. For this bot, I don't think learning is needed, as the information given almost never change for some period, it can be based on rules only.

**User 6**

I already try the system, I found that the system doesn't accept short phrases such as U instead of You, Also, When I send the QA there was no replay only the system said that your QA will be answered later like that. So,it is slowly

That are my comments,I hope it will help you

**User 7**

I tested the system and here are my observations:

 - It works well but only for very general cases. For instance I tried asking for application deadlines for an MSc in computer Science and I got the entry qualifications to the programme.

- I like the fact that if you say that your question was not answered you get a link that probably has the information you need.

That's all, Otherwise it is a great project.